\documentstyle[aps,epsf,prl,epsfig]{revtex}
\def\be{\begin{equation}}
\def\ee{\end{equation}}
\def\bea{\begin{eqnarray}}          
\def\eea{\end{eqnarray}}
\def\bi{\begin{itemize}}
\def\ei{\end{itemize}}

\begin{document}

\title{ Images of a Bose-Einstein condensate at finite temperature
\thanks{ Invited talk given at the conference ``Quantum Optics VI'', 
         June 13-18 2005, Krynica, Poland                             }  }

\author{ Jacek Dziarmaga }

\address{ Institute of Physics and Centre for Complex Systems,
          Jagiellonian University,
          Reymonta 4, 30-059 Krak\'ow, Poland  }

\date{ June 28, 2005 }

\maketitle

\begin{abstract}
A condensate initially prepared at finite temperature evolves under
external time-dependent perturbation into a time-dependent mixed state. 
In these notes I use number-conserving time-dependent Bogoliubov theory 
to derive probability distribution for different outcomes of density 
measurement on the time-dependent excited state. 
\end{abstract}

\section{Introduction}

Quantum measurements on Bose-condensed systems can give quite unexpected
results. For example, in the classic paper by Javanainen and Yoo
\cite{Fock} a density measurement on a Fock state $|N/2,N/2\rangle$ with
$N$ particles equally divided between two counter-propagating plane waves
$e^{\pm ix}$ reveals an interference pattern $\rho(x|\varphi)\sim
\cos^2(x-\varphi)$ with a phase $\varphi$ chosen randomly in every
realization of the experiment. The Fock state has a uniform single
particle density distribution, but its measurement unexpectedly reveals
interference between the two counter-propagating condensates. The Fock
state is a quantum superposition over $N$-particle condensates with
different relative phases $\varphi$ in their wave functions \cite{phase},
$|N/2,N/2\rangle\sim\int
d\varphi~|N:e^{+i(x-\varphi)}+e^{-i(x-\varphi)}\rangle$, but every single
realization of the experiment reveals such a density distribution as if
the state before the density measurement were one of the condensates
$|N:e^{+i(x-\varphi)}+e^{-i(x-\varphi)}\rangle$ with a randomly chosen
phase $\varphi$. This effect is best explained \cite{phase} when the
density measurement, which is a destructive measurement of all particle
positions at the same time, is replaced by an equivalent sequential
measurement of one position after another.  With an increasing number $n$
of measured positions a quantum state of the remaining $N-n$ particles
gradually ``collapses'' from the initial uniform superposition over all
phases to a state with a more and more localized phase $\varphi$.  For a
large $N$ a measurement of only a small fraction $\frac{n}{N}\ll 1$ of all
particles practically collapses the state of remaining $N-n$ particles to
a condensate with definite phase $\varphi$.

A lesson from this instructive example \cite{Fock,phase} is that quantum
measurement on an $N$-particle state with highly occupied single particle
modes ``collapses'' the state to a definite condensate with a definite
condensate wave function $\phi(x)$. The question is: what is the
probability distribution for different measurement outcomes $\phi(x)$? As
the set of condensates is not an orthonormal basis this is not a trivial
question.

In Ref.\cite{dynBog} we derived this probability distribution in the
framework of the time-dependent Bogoliubov theory at zero temperature. At
zero temperature a condensate initially prepared in its $N$-particle
ground state evolves under external time-dependent perturbation into a
time-dependent excited state. The excited state is a time-dependent
Bogoliubov vacuum i.e. at any time $t$ there exists a complete set of
quasiparticle annihilation operators for which the excited state is a
vacuum. In Ref.\cite{dynBog} it was shown that the time-dependent vacuum
has a simple diagonal structure which directly leads to a compact gaussian
probability distribution for different condensate wave functions
$\phi(x)$. As the case of zero temperature is covered in
Ref.\cite{dynBog}, in these notes I describe the general case of finite
temperature when the initial state is a condensate in equilibrium with a
thermal cloud of atoms. I derive gaussian probability distribution for
$\phi(x)$ at any time $t$ when the external perturbation drives the
initial thermal state into an excited mixed state.

\section{ N-conserving Bogoliubov theory }

Number conserving Bogoliubov theory \cite{NBT} is a quadratic 
approximation to the second quantized Hamiltonian (in trap units)
\be
\hat H~=~
\int dx~
\left[
\frac12 \partial_x\hat\Psi^\dagger \partial_x\hat\Psi+
\frac12 x^2 \hat\Psi^\dagger\hat\Psi+
V(t,x) \hat\Psi^\dagger\hat\Psi+
\frac12 g \hat\Psi^\dagger\hat\Psi^\dagger\hat\Psi\hat\Psi
\right]~.
\label{hatH}
\ee 
Here $\hat\Psi(x)$ is the bosonic annihilation operator, $V(t,x)$ is 
the external perturbation potential, and $g$ is strength of contact 
interaction between atoms. Here and in the following I will use 
one-dimensional notation but all equations can be generalized by the 
simple replacement $x\to \vec x$. The annihilation operator is split into 
condensate and non-condensate part
\be
\hat\Psi(x)~=
~\hat a_0~\phi_0(x)~+~\delta\hat\psi(x)~.
\label{hatPsi}
\ee
It is assumed that most atoms occupy the condensate mode $\phi_0(x)$.
Equation (\ref{hatPsi}) is substituted to the Hamiltonian (\ref{hatH}) and
then the Hamiltonian is expanded in powers of the fluctuation operator
$\delta\hat\psi$, see Ref.\cite{NBT}.

Many experiments on dilute atomic condensates can be clearly divided in
two steps: as a first step a condensate is prepared in its ground state
and then in the second step an external potential $V(t,x)$ is applied to
manipulate with the condensate wave function. Generic examples are phase
imprinting of dark solitons \cite{Hannover}, atomic interferometry 
\cite{interferometer}, or generation of shock waves in
Bose-Einstein condensates \cite{Bodzio}. At finite temperature the initial
state before the manipulation is a thermal state including thermal
excitations above the $N$-particle ground state. The initial condensate
wave function $\phi_0$ solves the stationary Gross-Pitaevskii equation
\be
\mu\phi_0=-\frac12\partial_x^2\phi_0+\frac12x^2\phi_0+g|\phi_0|^2\phi_0~.
\label{GP}
\ee
In Bogoliubov approximation the ground state is Bogoliubov vacuum 
$|0_b\rangle$ which can be written as a gaussian superposition over 
condensates \cite{vortex}
\be
|0_b\rangle ~=~
\int d^2b ~ 
e^{-\frac12\sum_{m=1}^M b_m^*b_m} ~
\left|
N:
\phi_0(x)+
\frac{1}{\sqrt{N}}
\sum_{m=1}^M
b_m u_m(x) + b_m^* v_m^*(x)
\right\rangle~.
\ee
Here the state $|N:\phi\rangle$ is a condensate of $N$ atoms in the normalized
condensate wave function $\frac{\phi}{\sqrt{\langle\phi|\phi\rangle}}$. The
Bogoliubov modes $u_m$ and $v_m$ are eigenmodes of the stationary 
Bogoliubov-de Gennes equations
\bea
\omega_m u_m &=&
-\frac12\partial_x^2u_m+
\frac12x^2u_m+
2g|\phi_0|^2u_m+
g\phi_0^2v_m~, 
\nonumber \\
-\omega_m v_m &=&
-\frac12\partial_x^2v_m+
\frac12x^2v_m+
2g|\phi_0|^2v_m+
g\left(\phi_0^*\right)^2u_m~. 
\label{BdG}
\eea 
Numerical solution of these equations gives a finite number of modes $M$.
At finite temperature the initial state is a thermal state $\hat\rho(0)$ 
with thermal quasiparticle excitations. The thermal state is also a 
gaussian state \cite{vortex}
\bea
\hat\rho(0) &=&
\int d^2b_L \int d^2b_R~  
e^{-\frac12b_L^*b_L-\frac12b_R^*b_R+b_L^*e^{-\beta\omega}b_R} ~
\nonumber\\
&&
\left|
N:
\phi_0(x)+
\frac{1}{\sqrt{N}}
\sum_{m=1}^M
b_{L,m} u_m(x) + b_{L,m}^* v_m^*(x)
\right\rangle
\left\langle
N:
\phi_0(x)+
\frac{1}{\sqrt{N}}
\sum_{m=1}^M
b_{R,m} u_m(x) + b_{R,m}^* v_m^*(x)
\right|~.
\label{hatrho}
\eea
Here $b_L^*b_L=\sum_{m=1}^Mb_{L,m}^*b_{L,m}$ and
$b_L^*e^{-\beta\omega}b_R=\sum_{m=1}^M b_{L,m}^*e^{-\beta\omega_m}b_{R,m}$.

In Bogoliubov theory the initial thermal state evolves under external
perturbation $V(t,x)$ into an excited state $\hat\rho(t)$ which has the
same form as the initial $\hat\rho(0)$ in Eq.(\ref{hatrho}) but with 
time-dependent Bogoliubov modes $u_m(t,x)$ and $v_m(t,x)$ which solve 
time-dependent Bogoliubov-de Gennes equations
\bea
i\partial_t u_m &=&
-\frac12\partial_x^2u_m+
\frac12x^2u_m+
2g|\phi_0|^2u_m+
g\phi_0^2v_m~, 
\nonumber \\
-i\partial_t v_m &=&
-\frac12\partial_x^2v_m+
\frac12x^2v_m+
2g|\phi_0|^2v_m+
g\left(\phi_0^*\right)^2u_m~ 
\label{tBdG}
\eea 
with initial conditions being the eigenmodes of the stationary BdG
equations (\ref{BdG}). The time-dependent condensate wave function 
$\phi_0(t,x)$ solves the time-dependent Gross-Pitaevskii equation
\be
i\partial_t\phi_0=
-\frac12\partial_x^2\phi_0+\frac12x^2\phi_0+g|\phi_0|^2\phi_0~.
\label{tGP}
\ee

\section{Probability distribution for outcomes}

As mentioned before, density measurement is ``collapsing'' $N$-particle
state to a Bose-Einstein condensate. The aim of the measurement theory is
to provide probability distribution for different condensate wave
functions $\phi$. In the present context of Bogoliubov theory it is 
convenient to split possible condensate wave functions into the condensate 
part and the non-condensate part:  
$\phi=\phi_0+\frac{1}{\sqrt{N}}\delta\phi$. The aim is to find
gaussian probability distribution for $\delta\phi$ in the gaussian state
$\hat\rho(t)$. Ideally the gaussian distribution would be fully determined
by the following equalities between second order correlators of the
gaussian $\delta\phi(x)$ and second order correlators of the field
operators:
\begin{eqnarray} 
\overline{\delta\phi^*(x) \delta\phi(y) }&\stackrel{?}{=}
\langle\delta\hat\psi^\dagger(x) \delta\hat\psi(y) 
\rangle=&
\sum_m n_m u_m^*(x)u_m(y)+
       (1+n_m) v_m(x) v_m^*(y)~, 
\label{1rho}\\
\overline{\delta\phi(x) \delta\phi^*(y) }&\stackrel{?}{=}
\langle 
\delta\hat\psi(x) \delta\hat\psi^\dagger(y) 
\rangle=&
\sum_m n_m v_m^*(x) v_m(y)+
       (1+n_m) u_m(x) u_m^*(y)~, \\
\overline{\delta\phi(x) \delta\phi(y) }&=
\langle 
\delta\hat\psi(x) \delta\hat\psi(y) 
\rangle=&
\sum_m n_m v_m^*(x) u_m(y)+
       (1+n_m) u_m(x) v_m^*(y)~, \\
\overline{\delta\phi^*(x) \delta\phi^*(y) }&=
\langle 
\delta\hat\psi^\dagger(x) \delta\hat\psi^\dagger(y) 
\rangle=&
\sum_m n_m u_m^*(x) v_m(y)+
       (1+n_m) v_m(x) u_m^*(y)~.
\label{corr}
\end{eqnarray}
Here the most right hand sides follow from the Bogoliubov theory
\cite{NBT}. $n_m=(e^{\beta\omega_m}-1)^{-1}$ is average number 
of thermally excited Bogoliubov quasiparticles in the initial state.
Unfortunately, because of the non-zero commutator
$[\delta\hat\psi(x),\delta\hat\psi^\dagger(y)]=\delta(x-y)-\phi_0^*(x)\phi_0(y)$,
the first two conditions cannot be satisfied simultaneously. I replace
them with the condition
\begin{eqnarray}
\overline{\delta\phi^*(x) \delta\phi(y) } ~=~
\left(\overline{\delta\phi(x)\delta\phi^*(y)}\right)^* ~=~
\frac12
\langle
\delta\hat\psi^\dagger(x) \delta\hat\psi(y)+
\delta\hat\psi(y) \delta\hat\psi^\dagger(x) 
\rangle~.
\label{replacement}
\end{eqnarray}
It is convenient to expand the fluctuation as
\be
\delta\phi(x|z)=
\sum_{\alpha=1}^\infty z_\alpha \phi_\alpha(x)
\ee
in the ortonormal basis of the eigenmodes $\phi_\alpha$ of the reduced
single particle density matrix
\be
\langle \delta\hat\psi^\dagger(x) \delta\hat\psi(y) \rangle=
\sum_{\alpha=1}^M \delta N_\alpha~ \phi_\alpha^*(x)\phi_\alpha(y)~.
\label{1PDM}
\ee
Here the left hand side is given by Eq.(\ref{1rho}). The right hand side
is obtained after diagonalization of the hermitean operator on the left.
The real eigenvalues $\delta N_\alpha$ are average occupation numbers of
the corresponding non-condensate modes $\phi_\alpha$. The correlators
(\ref{corr}) after the semiclassical approximation (\ref{replacement})
determine the matrix of correlators of the complex gaussian random 
variables $z_\alpha$:
\be
\left(
\begin{array}{cc}
\overline{z_\alpha^\star z_\beta}        & 
\overline{z_\alpha^\star z_\beta^\star}  \\
\overline{z_\alpha z_\beta}              & 
\overline{z_\alpha z_\beta^\star}        \\
\end{array}
\right)=
\left(
\begin{array}{cc}
D_{\alpha\beta} &
C_{\alpha\beta} \\
C^*_{\alpha\beta} &
D_{\alpha\beta} \\ 
\end{array}
\right)~.
\label{zcorr}
\ee
Here the $M\times M$ matrices on the right hand side are
\bea
D_{\alpha\beta}&=&
\frac12
\left(
U^*_{\alpha m}n_mU_{\beta m}+V^*_{\alpha m}(1+n_m)V_{\beta m}+
U_{\alpha m}(1+n_m)U^*_{\beta m}+V_{\alpha m}n_mV^*_{\beta m}
\right)~,
\label{D}\\
C_{\alpha\beta}&=&
U^*_{\alpha m}n_mV^*_{\beta m}+V^*_{\alpha m}(1+n_m)U^*_{\beta m}~,
\label{C}
\eea
with the matrix elements $U_{\alpha m}=\langle\phi_\alpha|u_m \rangle$ and
$V_{\alpha m}=\langle\phi_\alpha|v^*_m\rangle$. Replacing $z_\alpha$'s
with real coordinates, $z_\alpha=x_\alpha+iy_\alpha$, we get a real
symmetric matrix of correlators
\be
\left(
\begin{array}{cc}
\overline{x_\alpha x_\beta} & 
\overline{x_\alpha y_\beta} \\
\overline{y_\alpha x_\beta} &
\overline{y_\alpha y_\beta} \\
\end{array}
\right)=
\frac12
\left(
\begin{array}{cc}
{\rm Re }~ D_{\alpha\beta}+
{\rm Re }~ C_{\alpha\beta} &
{\rm Im }~ C_{\alpha\beta} \\
{\rm Im }~ C_{\alpha\beta} &
{\rm Re }~ D_{\alpha\beta}-
{\rm Re }~ C_{\alpha\beta} \\
\end{array}
\right)
\label{xx}
\ee  
and a condition that ${\rm Im }~D_{\alpha\beta}=0$ - a good test of the
correctness of the calculations. Diagonalization of the correlation matrix
(\ref{xx}) gives eigenvalues $\lambda_s\geq 0$ with $s=1,..,2M$.  
Corresponding eigenvectors are columns of an orthogonal matrix $O$. The
eigenvectors define convenient parametrization of the gaussian fluctuation
as
\be
\delta\phi(x)=
\sum_{\alpha=1}^M
z_\alpha
\phi_{\alpha}(x)=
\sum_{\alpha=1}^M
\phi_\alpha(x)
\sum_{s=1}^{2M}
\left(O_{\alpha,s}+iO_{M+\alpha,s}\right)q_s\equiv
\sum_{s=1}^{2M} \Phi_s(x) q_s~
\label{zq}
\ee
with independent real gaussian random variables $q_s$ of zero mean 
and variances $\overline{q_s^2}=\lambda_s$. However, this is not the end
of the story yet.

As a result of the semiclassical approximation in Eq.(\ref{replacement})
averages like e.g. average density of depletion
$\overline{\delta\phi^*(x)\delta\phi(x)}$ are divergent because there is
infinite number of unoccupied modes $\phi_\alpha(x)$, every one of them
contributing to this depletion density a term 
$\frac12\phi_\alpha^*(x)\phi_\alpha(x)$. In stochastic averages like 
$\overline{\delta\phi^*(x)\delta\phi(x)}$ average occupation numbers
of modes $\phi_\alpha(x)$ seem to be $\delta N_\alpha+\frac12$ instead of
the correct $\delta N_\alpha$. This artifact of the semiclassical 
approximation can be corrected by introducing to Eq.(\ref{zq}) of 
regularization factors:
\be
\left.\delta\phi(x)\right|_{\rm reg}=
\sum_{\alpha=1}^M
z_\alpha
\left(
\frac{\delta N_\alpha}{\delta N_\alpha+\frac12}
\right)^{1/2}
\phi_{\alpha}(x)=
\sum_{\alpha=1}^M
\left(
\frac{\delta N_\alpha}{\delta N_\alpha+\frac12}
\right)^{1/2}
\phi_\alpha(x)
\sum_{s=1}^{2M}
\left(O_{\alpha,s}+iO_{M+\alpha,s}\right)q_s\equiv
\sum_{s=1}^{2M} \Phi_s(x) q_s~
\label{zqreg}
\ee
As expected in semiclassical approximation, the regularizing factors
$\left(\frac{\delta N}{\delta N+\frac12}\right)^{1/2}$ are approximately
$1$ for the highly occupied modes with $\delta N_\alpha\gg 1$ which
dominate in the density distribution, but at the same time they remove
the divergence coming from the infinity of unoccupied modes. The wave
functions $\Phi_s(x)$ are in general neither normalized nor orthogonal,
except in the quantum limit of zero temperature, see the proof in
Ref.\cite{dynBog}.

\section{ Conclusion }

In conclusion, a recipe to simulate density measurement on the 
time-dependent excited thermal state has the following steps:

\bi 

\item Solve stationary Gross-Pitaevskii and Bogoliubov-de Gennes equations
(\ref{GP},\ref{BdG}) to provide initial conditions for $\phi_0(t,x)$,
$u_m(t,x)$ and $v_m(t,x)$, and the initial quasiparticle frequencies
$\omega_m$.

\item Solve time-dependent Gross-Pitaevskii and Bogoliubov-de Gennes
equations (\ref{tGP},\ref{tBdG}) with respect to $\phi_0(t,x)$,
$u_m(t,x)$, and $v_m(t,x)$.

\item Diagonalize the reduced single particle matrix (\ref{1PDM}) to get
its non-condensate eigenmodes $\phi_\alpha$ with their average occupation
numbers $\delta N_\alpha$.

\item Build the matrices $D_{\alpha\beta}$ and $C_{\alpha\beta}$ in
Eqs.(\ref{D},\ref{C}), and then the real symmetric correlation matrix in
Eq.(\ref{xx}).

\item Diagonalize the correlation matrix in Eq.(\ref{xx}) to get its real
eigenvalues $\lambda_s$ and correposnding eigenvectors $O_{\alpha s}$.

\item Build the regularized modes $\Phi_s$ according to their definition 
implicit in Eq.(\ref{zqreg}):
\be
\Phi_s(x)~=~
\sum_{\alpha=1}^M
\left(\frac{\delta N_\alpha}{\delta N_\alpha+\frac12}\right)^{1/2}
\phi_\alpha(x)~ 
\left( O_{\alpha,s}+iO_{M+\alpha,s} \right)~.
\ee
   
\item Choose independent real random variables $q_s$'s from 
their gaussian distributions 
of zero mean and variance $\overline{q_s^2}~=~\lambda_s$, and then 
combine the chosen $q$'s into condensate density 
\be
\rho(x|q)~=~
\left|
\sqrt{N}\phi_0(x)~+~
\sum_{s=1}^{2M} q_s \Phi_s(x)
\right|^2~.
\ee

\ei
The $\rho(x|q)$ defines a family of all possible density measurement 
outcomes with a gaussian probability distribution for different $q$'s. 

In the limit of zero temperature this general recipe coincides with 
the recipe derived by different methods in Ref.\cite{dynBog}. At zero 
temperature the wave functions become $\Phi_\alpha(x)\sim\phi_\alpha(x)$ 
for $\alpha=1,...,M$ and zero otherwise (here the $\sim$ means equality
up to a phase factor). Corresponding variances are 
$\overline{q_\alpha^2}=\delta N_\alpha$.

\section*{ Acknowledgements } 

I would like to thank Zbyszek Karkuszewski and Krzysztof Sacha for
stimulating discussions. This work was supported in part by Polish
government scientific funds (2005-2008) as a research project.



\end{document}